\documentclass[amssymb,aps,showpacs,nofootinbib,twocolumn,letterpaper]{revtex4}
\usepackage[]{graphicx}
\usepackage[]{amsmath}
\usepackage{hyperref}

\DeclareMathOperator{\Tr}{Tr}
\DeclareMathOperator{\arctanh}{arctanh}

\def\b#1{\mathbf{#1}}

\def\12{{\textstyle \frac{1}{2}}}
\def\32{{\textstyle \frac{3}{2}}}
\newcommand{\proj}[1]{\ket{#1}\bra{#1}}

\def\sym{\text{\rm sym}}

\def\ket#1{| #1 \rangle}
\def\bra#1{\langle #1 |}

\def\vev#1{\langle #1 \rangle}

\def\cE{\mathcal{E}}

\def\bC{\mathbb{C}}

\begin{document}

\title{Dynamics of a Quantum Reference Frame}
\author{David Poulin}
\email{dpoulin@ist.caltech.edu}
\affiliation{Center for the Physics of Information, California Institute of Technology, Pasadena, CA 91125}
\author{Jon Yard}
\affiliation{Institute for Quantum Information, California Institute of Technology, Pasadena, CA 91125}

\date{\today}

\begin{abstract}
We analyze a quantum mechanical gyroscope which is modeled as a large
spin and used as a reference against which to measure the angular
momenta of spin-1/2 particles.  These measurements induce a
back-action on the reference which is the central focus of our study.
We begin by deriving explicit expressions for the quantum channel
representing the back-action. Then, we analyze the dynamics
incurred by the reference when it is used to sequentially measure
particles drawn from a fixed ensemble.  We prove that the reference
thermalizes with the measured particles and find that generically, the
thermal state is reached in time which scales linearly with the size
of the reference.  This contrasts a recent conclusion of Bartlett et
al.\ that this takes a quadratic amount of time when the particles are
completely unpolarized. We  now understand their result in terms of a
simple physical principle based on symmetries and conservation laws.
Finally, we initiate the study of the non-equilibrium dynamics of the
reference.  Here we find that a reference in a coherent state will
essentially remain in one when measuring polarized particles, while
rotating itself to ultimately align with the polarization of the
particles.  
\end{abstract}

\maketitle

\section{Introduction}

Physical systems are described in terms of kinematical observables such as position, momentum, and spin. In the presence of symmetries, only a subset of those quantities -- the physical observables --  are physically relevant, as they too must obey the same symmetries.  One way of obtaining physical observables is to use some systems as references relative to which the properties of other systems can be measured, yielding observables which are \emph{relational}. The state of the reference on its own cannot be measured: only its correlations with other systems acquire physical meaning through direct measurements. In many circumstances, the existence of a suitable reference is conspicuous, often blurring the distinction between kinematical and relational observables. 

For example, the walls of a laboratory can be used to align a Stern-Gerlach apparatus to very high precision. This reference enables one to measure any component of an electron's spin, despite the rotational invariance of the underlying theory. While this statement may seem contradictory, it is only due to the fact that the reference was omitted from the description of the measured quantity. Formally, the spin of the electron is only determined {\em relative} to the orientation of the measurement apparatus (and ultimately the laboratory walls, the stars, etc.) and all symmetries are respected. When the reference systems are classical, whether or not to include them in a given physical description is purely an aesthetic choice which has no impact on physical predictions \cite{BRS05a}. These ``two sides of the same coin" are analogous to doing calculations in electromagnetism with a special choice of gauge or using the gauge invariant language: each approach has its advantages. 

However, if the measuring device (or at least certain essential parts of it) are inherently \emph{quantum mechanical}, these distinctions become more relevant.  For example, it was pointed out long ago by Wigner \cite{Wig52a} that a quantum mechanical conservation law places constraints on the ability to obtain precise measurements of certain observables (see also the strengthening \cite{AY60a}).  Similar arguments can be made for clocks, rods and gyroscopes, as they are also often taken for granted and omitted from the theoretical description of the situation at hand.

In this paper, we study a model of quantum measurement in which the task is to measure the angular momentum of spin-$\frac 12$ particles along some given direction, in the presence of a rotational symmetry.   The information about the axis along which to measure is supplied by a quantum mechanical reference system.  We will not only study the limitations this model places on performing perfect measurements -- rather, we will focus on how the measurements affect the reference, and thus how they change the character of measurements performed with the same reference at later times. 

While omitting the description of the reference at hand is often entirely justified from a practical standpoint, this issue has been the source of a great deal of confusion and heated debates. One illustration, particularly disputed in the Bose-Einstein Condensation (BEC) community, is the definition of an order parameter as either a local expectation or a correlation function. The former idea leads to the concept of spontaneous symmetry breaking \cite{And72a}, where the field acquires a finite expectation value prohibited by the symmetry. The latter idea views one part of the BEC as a reference relative to which the expectation of the field at another point in the BEC can be measured, thus illustrating the importance of correlation functions \cite{Leg01a}. Similar debates have taken place with regard to the phase of a laser field \cite{Mol97a} being either absolute or only defined relative to a reference pulse. 

Another important example occurs in general relativity, where the group of symmetries consists of diffeomorphisms of spacetime. Indeed, the issues concerning physical observables in general relativity are among the most profound in physics as they raise, among other things, questions regarding the measurability of time \cite{AB61a,PW83b}, as well as that of the spacetime metric \cite{Ber61a} itself.  Once again, some of these conceptual difficulties can be circumvented by a careful analysis of the role of the reference systems involved in our description of nature \cite{PW83b,AK84a,Rov91a}.   We expect that  resolving such foundational issues would be an important step towards the much-sought-after marriage between quantum mechanics and general relativity
at the heart of current theoretical physics research.

Quantum mechanical descriptions of reference systems have revealed important physical consequences.  These include the possibility of breaking certain superselection rules \cite{AS67a,KMP04a} and the existence of an intrinsic decoherence in quantum gravity \cite{Mil91a,GPP04a,Pou06a,MP06a}. However, it is the study of quantum reference systems as {\em physical resources} that has recently received the most attention. Indeed, using the tools of quantum information science, one can quantify the cost of establishing a reference between two distant parties \cite{PS01a}, the cost of realizing a reference with a desired degree of accuracy \cite{BRS04a}, etc.\ (see \cite{BRS06a} for a recent review). 

The latest addition to the theory of quantum reference systems is the study of their {\em degradation} under repeated measurements \cite{BRST06a,BRST06b}. Indeed, the very act of using a reference to measure a system typically decreases the accuracy with which it can make a subsequent measurement on another system. Studying specific models, the authors of \cite{BRST06a,BRST06b} came to the conclusion that this measurement accuracy typically decreases as the square-root $\sqrt t$ of the number of measured systems $t$, or in other words, that the so-called longevity of a reference is proportional to the square of its size. Here, we revisit one of their models and analyze its dynamics in more detail. We show how their result follows from simple physical principles based on symmetries and conservation laws which are applicable even to general systems.  We also find that the case analyzed by those authors is in fact quite singular and does not reflect the general behavior of a reference's longevity, which we instead find to scale {\em linearly} with its size in the generic case.  As we observe in Sec.~\ref{sec:timeA}, this distinction exists because the source particles cause the reference to undergo a random walk whose dynamics are dominated by fluctuations when the source particles are unpolarized and by drift when they are polarized. 

The importance of studying quantum reference frame degradation stems not only from its repercussions on fundamental physics outlined above, but also from its relevance to realistic experimental settings (see e.g.\ \cite{WDL+04a,RBMC04a} for a description of experimental realizations relatively similar to our theoretical idealization). In particular, such effects are likely to play an important role in the realization of quantum information processors. We will elaborate on this aspect in Sec.~\ref{sec:discussion}. 

The rest of this paper is organized as follows. Sec.~\ref{sec:model} sets the stage by introducing the model that we study and describing the task it is meant to accomplish.  Sec.~\ref{sec:asymptotic} presents an analysis of the asymptotic behavior of the reference system. The following section studies the thermalization time required to reach this asymptotic state. In Sec.~\ref{sec:non-eq.}, we derive semi-classical equations of motion that model the dynamics of the reference system during the equilibration process.  The last section summarizes our results and discuss some of their implications.

\section{The model}
\label{sec:model}
Our primary object of study is essentially a gyroscope, i.e.\  a physical system that singles out a particular direction in three-dimensional space.   
%A DR can be used to construct constrained relational observables in a rotationally invariant theory -- that is, one which is invariant under the rotation group $SU(2)$.  
In principle, an ideal (classical) gyroscope would allow an experimenter to measure the spin of a spin-$\frac 12$ \emph{source} particle along the axis of rotation of the gyroscope.  The quantum analog of a gyroscope is a system with a large amount of spin; we henceforth refer to such a system as a \emph{reference}.  The state of the reference is described by a density matrix $\rho$ on the spin-$\ell$ irreducible representation (irrep) of $SU(2)$.  Throughout, we will consider $\ell$, or equivalently the dimension $d=2\ell + 1$, as an indication of the \emph{size} of the reference.

Ideally, the reference will be in a coherent state and will thus have a maximal amount of angular momentum concentrated in some direction $\hat{\b{n}}$.  We only allow the experimenter to operate on the system and the reference in a rotationally invariant manner. Thus, the only non-trivial observable to which the experimenter has access is the total angular momentum of the source particle and reference.  When used in this fashion, the quantum reference differs from its classical counterpart in two fundamental ways: 
\begin{enumerate}
\item The measurements will only be an approximation of what would be obtained by using the corresponding classical reference.
\item Each time the reference is used, it suffers an inevitable ``back-action" which ultimately changes the character of future measurements.
\end{enumerate}

In any practical scenario, this reference would in general consist of a {\em composite} system, e.g.\ composed of  the many electron spins of a ferromagnet. That only the $SU(2)$ degrees of freedom represented by $\rho$  are of interest implies that we are assuming that all the constituent particles of the gyroscope are used in a symmetric manner, and so can be considered as a single particle with large angular spin $\ell$, such as a Rydberg atom. Other interactions will typically lead to a shorter longevity of the reference, but we are here interested in the fundamental upper bound which physics places on its longevity. 

While the physics we analyze is independent of any background reference frame, our mathematical analysis requires us to choose one.  This artificial background coordinate frame should be thought of as a choice of gauge with respect to which the underlying physics is insensitive.  In terms of these background coordinates, the $SU(2)$ generators of the reference are kinematic observables which we denote $\b{L} = (L_x,L_y,L_z)$.  These operators satisfy (see e.g.\ \cite{Sak94a})
\begin{equation}
[L_i,L_j] = i\epsilon^{ijk}L_k
\label{eq:su2}
\end{equation}
where $\epsilon$ is the completely antisymmetric tensor for which $\epsilon^{xyz} = 1$.  The corresponding raising and lowering operators $L_{\pm} = L_x \pm i L_y$ satisfy 
\begin{equation}
[L_+,L_-] = 2 L_z, \hspace{.3in} [L_z,L_\pm] = \pm L_\pm.
\label{eq:ladder}
\end{equation}

Consider now a source which emits spin-$\12$ particles, each described by the same density matrix $\xi$. The corresponding angular momentum operators $\b{S} = (S_x, S_y,S_z)$ also satisfy the commutation relations given by Eq.~(\ref{eq:su2}). The constrained relational observable allowing us to measure the spin of the particle relative to the quantum reference is the square of the total spin operator $\bf{J}^2 = (\bf{L}+\bf{S})^2$. Since $\bf{L}^2$ and $\bf{S}^2$ are in the center of the algebra, $\bf{J}^2$ (and arbitrary functions thereof) is the only non-trivial operator that can be measured in our toy universe. Measurement of this observable projects the combined state of the system and the reference onto an eigenspace of $\b{J}^2$ with eigenvalue $j(j+1)$, where $j = \ell \pm \frac{1}{2}$.  After this measurement, we assume that the measured source particle is discarded and that the measurement result is forgotten; in other words, the measurement scheme is non-adaptive.  While it is an interesting question whether or not the history of results from subsequent measurements could potentially be used to give a more accurate description of the dynamics of the reference, we do not consider such approaches in this paper.

\subsection{The approximate measurement}

We now describe the measurement induced on the source particle by the joint measurement of $\bf{J}^2$. This projective measurement on the combined system induces a positive operator valued measurement (POVM) on the source.  To find an expression for this POVM, we begin by writing the respective projections onto the $j=\ell\pm\12$ irreps as
 \begin{equation}
 \Pi_\pm = \frac 12\left(I_{2d} \pm \frac{4\b{L}\cdot\b{S} + I_{2d}}{
d}\right).
 \label{projections}
 \end{equation}
The measurement $\{\Lambda^\pm_\rho\}$ induced on the source is now given by a partial trace over the reference, which yields
\begin{eqnarray*}
\Lambda^\pm_\rho &=& \Tr_{R}\Pi_+(\rho\otimes I_2) = \frac 12\left(I_2 \pm \frac{4\vev{\b{L}}\cdot \b{S} + I_2}{d}\right).  %\\
%\Lambda^-_\rho &=& \Tr_{RF}\Pi_-(\rho\otimes I_2) = \frac{1}{d}\big(\ell -  2\vev{\b{L}}\cdot\b{S}\big).  
\end{eqnarray*}
Note that the measurement depends only on the expectations $\vev{L_i} \equiv \Tr\rho L_i$ of the angular momentum of the reference.  Defining the 3-vector ${\bf n}_\rho = \vev{{\bf L}}/(\ell + \frac 12)$, whose norm is bounded by 1, this becomes
\begin{eqnarray}
\Lambda^+_\rho &=& \frac{\ell+1}{d}I_2  +  {\bf n}_\rho \cdot \bf{S} \label{eq:Lp}\\
\Lambda^-_\rho &=&  \frac{\ell}{d}I_2 -  {\bf n}_\rho \cdot \bf{S} . \label{eq:Lm} 
\end{eqnarray}
Thus, we see that the relational observable is approximating a measurement of the spin of the source particle along the axis $\hat{\bf n}_\rho = \vev{{\bf L}}/|\vev{{\bf L}}|$, or rather, of the observable $\hat{\bf  n}_\rho \cdot {\bf S}$. In the classical limit, where $|\vev{\bf{L}}| \rightarrow \ell \rightarrow \infty$, the POVM elements become the spectral projectors associated to the observable $\hat{\bf n}_\rho \cdot \bf{S}$, and so the measurement is perfect.

\subsection{An operational criterion for the quality of the measurement}
\label{sec:merit}
Suppose that we set out to measure the the spin of the source along the direction $\hat{\bf  n}$.  In other words, we seek to measure the projections 
\begin{equation}
P^\pm_{\hat n} = I_2/2 \pm \hat{\bf n} \cdot \bf{S}.
\label{eq:perfect}
\end{equation} 
As we have seen, the measurement $\{\Lambda^\pm_\rho\}$ which will actually be performed by our method will rather be an imperfect simulation of $\{P^\pm_{\hat n}\}$.  There are various operational criteria which can be defined to judge the quality of this simulation.  One such measure \cite{BRST06a} is the probability of correctly identifying a maximally polarized source which is prepared along the directions $\pm\hat{\b{n}}$ with equal probabilities.  This average success probability depends linearly on the inner product $\hat{\bf n} \cdot {\bf n}_\rho$ and can be written as 
\begin{equation*}
Q_\text{ave} = \frac 12\big(1 +  \hat{\bf n} \cdot {\bf n}_\rho\big).
\end{equation*}
Other figures of merit could be considered, such as the worst-case success probability for identfying the above source states, or perhaps another average error in which those source states appear with unequal probabilities.  It is however easy to see that these criteria differ from $Q_\text{ave}$ by $O(1/d)$ and are thus essentially equivalent. 
%Another figure of merit is the worst-case success probability of identifying the above mentioned source states. We will not commit to any particular operational figure of merit here since it is clear by comparing Eqs.~(\ref{eq:Lp},\ref{eq:Lm}) with Eq.~(\ref{eq:perfect}) that the distinction between the ideal projectors and the POVM elements induced by the relational observable is dictated by $\hat{\bf n} \cdot {\bf n}_\rho$. Any reasonable figure of merit should approach one as  $\hat{\bf n} \cdot {\bf n}_\rho \rightarrow 1$. 

There are two distinct properties of the reference that contribute to $\hat{\bf n} \cdot {\bf n}_\rho$. On the one hand, the length of ${\bf n}_\rho$ could be less than 1.  Indeed, when $\hat{\bf n}$ and ${\bf n}_\rho$ are aligned, we find that $\hat{\bf n} \cdot {\bf n}_\rho = |{\bf n}_\rho|$.  This is the scenario that was studied in \cite{BRST06a} and leads to a ``fuzzy" measurement --- a mixing of the $+$ and $-$ measurement outcomes.  On the other hand, $\hat{\bf n}$ and ${\bf n}_\rho$ may be misaligned, where the reference singles out a direction that differs from the direction we initially set out to measure. In this case, the reference still leads to precise measurements; they just don't happen to coincide with what we are interested in. An important result that we will establish here is that degradation of a reference under repeated uses typically lead to misalignment rather than fuzzy measurements.

The authors of \cite{BRST06a} defined the \emph{longevity} of the reference to equal the number of times that the reference can be used before the above figure of merit drops below some constant value.  They found that the longevity of a reference, when used to measure unpolarized source particles, scales quadratically with the size of the reference.  We will see that even if the source particles are only slightly polarized, a sufficiently large reference will typically have a longevity which scales \emph{linearly} with its size. 

\subsection{A quantum channel describing the back-action on the reference}

In this section we derive exact expressions for the disturbance experienced by the reference after being used to measure a source particle.  We will find that just as the induced POVM on the source particle depends only on the average angular momentum $\vev{\b{L}}$ of the reference, this back-action can be expressed as a \emph{quantum channel}, or completely positive trace preserving (CPTP) map, which depends only on the average angular momentum $\vev{\b{S}}$ of the source. Since
\[\b{J}^2 = (\ell+\12)(\ell + \32)\Pi_+ + (\ell-\12)(\ell+\12)\Pi_-,\]
a measurement of the total angular momentum is equivalent to the two-outcome von Neumann measurement $\{\Pi_+,\Pi_-\}$. 
Suppose the reference and source begin in the respective states $\rho$ and $\xi$, after which the above joint measurement is performed and the source particle is discarded. This process causes the reference state to undergo a discrete time evolution given by the CPTP map
\begin{equation*}
\cE_\xi(\rho) \equiv \Tr_S\big(\Pi_+(\rho\otimes \xi)\Pi_+ + \Pi_-(\rho \otimes \xi)\Pi_-\big).
\label{cptp}
\end{equation*}
Using (\ref{projections}), we may express this channel as 
\begin{eqnarray}
\cE_\xi(\rho) &=& \left(\frac{1}{2} + \frac{1}{2d^2}\right)\rho \nonumber \\
& & + \frac{8}{d^2}\Tr_S(\b{L}\cdot\b{S})(\rho\otimes\xi)(\b{L}\cdot\b{S}) 
 \label{eq:channel1}\\
& & + \frac{2}{d^2}\big((\b{L}\cdot\vev{\b{S}}) \rho + \rho(\b{L}\cdot\vev{\b{S}})\big).
\nonumber
\end{eqnarray}
It is clear that this expression is independent of the particular choice of background coordinates.   For a polarized source,  
we choose our background coordinate system so that $\xi$ commutes with $S_z$, allowing us to write $\xi = I/2 +2\vev{S_z}S_z$ and $\vev{S_x} = \vev{S_y} = 0$.
In this case, Eq.~(\ref{eq:channel1}) can be written
\begin{eqnarray*}
\cE_{\xi}(\rho) 
&=&  \left(\frac 12+\frac{1}{2d^2}\right)\rho + \frac{2}{d^2}\sum_{i=x,y,z} L_i \rho L_i
\\ & & + \frac{2\vev{S_z}}{d^2}\big(L_z\rho + \rho L_z  + L_+\rho L_- - L_-\rho L_+\big).
\end{eqnarray*} 
While not immediately clear from this expression, $\cE_\xi$ is invariant under rotations about the $z$ axis because the only symmetry breaking in (\ref{eq:channel1}) comes from the polarization of the source particles.  In the event that the source particle is completely unpolarized, this implies that the action of the channel $\cE_\xi$ is rotationally invariant. 
In this case, $\cE_\xi$ reduces to an instance of the so-called ``spin-$\ell$" channel introduced in \cite{Rit05a}.
A more useful expression for $\cE_\xi$  can be obtained by straightforward manipulations which lead to
\begin{eqnarray}
\cE_\xi(\rho) &=& \left(\frac 12 + \frac{1-4\vev{S_z}^2}{2d^2}\right)\rho \label{alternateform}\\
& & + \frac{2}{d^2}\big(L_z + \vev{S_z}\big)\rho\big(L_z + \vev{S_z}\big) \nonumber\\
& & + \frac{1+ 2\vev{S_z}}{d^2}L_+\rho L_- + \frac{1- 2\vev{S_z}}{d^2}L_-\rho L_+.\nonumber
\end{eqnarray}

When the reference is used to consecutively measure $t$ source particles, each of which is in the state $\xi$, it will experience a disturbance which is described by the channel $\cE_\xi^t = \cE_\xi \circ \cdots \circ \cE_\xi$ where $\circ$ denotes the usual composition. We may think of $t$ as a natural (discrete) time parameter for our problem and write the state of the reference after it has been used to measure $t$ particles as $\rho(t) = \cE^t_\xi(\rho)$. These dynamics are the central interest of this paper. 

The dynamics of the reference implies a time dependence of its angular momentum $\vev{{\bf L}(t)} = \Tr [\rho(t){\bf L}]$. Note that because the total angular momentum is a constrained relational observable, the measurement process preserves angular momentum. It may however create a flow of angular momentum from the reference to the source particles. This explains the time dependence of  $\vev{{\bf L}(t)}$. This time dependence will in turn induce a time dependence of the actual POVM $\big\{\Lambda^\pm_{\rho(t)}\big\}$ which is performed on the source particle at time step $t$ via Eqs.~(\ref{eq:Lp},\ref{eq:Lm}). Thus, the quality of the measurements obtained by repeatedly using the same reference may in general {\em degrade} over time. 

\section{Asymptotic behavior}
\label{sec:asymptotic}

We will now analyze the asymptotic state of the reference as $t\rightarrow \infty$. This is an important first step in understanding the general dynamics of a reference. In particular, if the asymptotic state of the reference induces good measurements of the source, we may conclude that it has an infinite longevity. 

\subsection{Thermodynamical approach}

We may think of the stream of source particles as an infinite bath with which the reference interacts. The von Neumann entropy of the bath particles is given by 
\[H(\xi) = H\big(\12 + \vev{S_z}\big)\]
where for scalar $p$, we have written 
\[H(p) = -p\log p -(1-p)\log(1-p)\] for the binary Shannon entropy. We may assume that the polarization of the source particles is induced by an external magnetic field of strength $B$ which is aligned with the positive $z$ axis.  
In this case, the average energy of each bath particle is equal to  $E = -B\vev{S_z}$.  In units where $B=1$, we therefore conclude that the bath is at inverse temperature
\begin{equation}
\beta = \frac{d H}{dE} = \log\left(\frac{1+2\vev{S_z}}{1-2\vev{S_z}}\right) = 2\arctanh(2\vev{S_z}). \label{eqn:beta}
\end{equation}
On physical grounds, we expect the reference to thermalize with the bath after being used to measure a large number of particles.  The asymptotic partition function is thus anticipated to be 
\begin{equation} 
Z =  \Tr\exp(-\beta L_z) = \frac{\sinh(\beta d/2)}{\sinh(\beta/2)},  \label{eqn:partition}
\end{equation}
leading to the asymptotic reference state $\rho_\infty = \frac 1Z \exp(-\beta L_z)$.
In the next subsection, we prove that this corresponds to the asymptotic state by showing that it is the unique fixed point of the quantum channel $\cE_\xi$.

Meanwhile, let us investigate the usefulness of this asymptotic state to serve as a reference. Clearly, the angular momentum $\vev{\bf L}$ of the reference in its asymptotic state is along the $z$ direction. Thus, if the original purpose of the reference was to measure the spin of the source particles along an axis that differs significantly from $z$, we conclude that it will fail to do so asymptotically simply because it will be misaligned.

\begin{figure}[t!]
{\hspace{-.25in}
\includegraphics[width=3.5in]{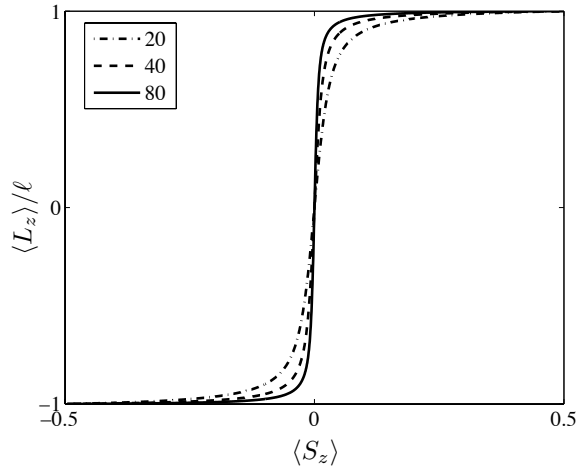}} 
\caption{Normalized polarization of the reference $\langle L_z \rangle/\ell$ thermalized with a source of polarization $\langle S_z \rangle$ for the various reference sizes $\ell = 20$, 40, and 80. The transition sharpens as the reference's size increases, with a width of roughly $1/\ell$, and to first order, we find that the slope of $\vev{L_z}$ is either $\ell$ or $1/\ell$ accordingly. In the macroscopic limit, any source polarization induces a full reference polarization.}
\label{fig:thermal}
\end{figure}

On the other hand, if the reference is meant to measure the spin of the source particles {\em along their axis of polarization}, then it may continue to serve this purpose in its asymptotic state. From our discussion in Sec.~\ref{sec:merit}, we know that the figure of merit is in this case given by the length of the vector ${\bf n_\rho} = \vev{{\bf L}}/(\ell+\frac 12)$.  The expectation of the $z$ component $\vev{L_z}$ of the asymptotic reference's angular momentum, as a function of the inverse temperature $\beta$, is given by 
\begin{eqnarray*}
\vev{L_z} &=& \frac{1}{Z}\frac{\partial Z}{\partial\beta} \\
&=& \frac{\sinh(\beta /2)}{\sinh(\beta d/2)} \frac{\partial}{\partial \beta}
\frac{\sinh(\beta d/2)}{\sinh(\beta /2)} \\
&=& \big(d\coth(\beta d/2) - \coth(\beta/2)\big)/2. 
\end{eqnarray*}
Using (\ref{eqn:beta}), we may rewrite this expression in terms of the source polarization $\vev{S_z}$  to obtain
\begin{equation*}
\vev{L_z} = \frac{d}{2} \coth\big(d\arctanh(2\vev{S_z})\big) - \frac{1}{4\vev{S_z}}.
\end{equation*}
Note that this expression is an antisymmetric function of $\vev{S_z}$.  In particular, the asymptotic reference state is well-defined when $\vev{S_z} = 0$ by continuity.  In Fig.~\ref{fig:thermal}, we have plotted the normalized quantity $\vev{L_z}/\ell$ for various reference sizes $\ell$.  From this, we can compute the figure of merit $|{\bf n_\rho}|$.  Observe that the case of an unpolarized source (i.e.\ $\vev{S_z} \approx 0$) is quite singular and leads to a completely different physical behavior of the reference than in the generic case. For the case of a vanishing source polarization in which $\vev{S_z}  \ll 1/\ell$, the figure of merit vanishes:
\begin{equation*}
|{\bf n_\rho}| = \frac 43 \ell\vev{S_z} + O\left(\frac{\vev{S_z}}{\ell}\right)^3.
\end{equation*}
On the other hand, for a source with a finite amount of polarization, where $\vev{S_z} \geq \ell^{-a}$ with $0<a<1$, we instead find that the figure of merit approaches 1:
\begin{equation*}
|{\bf n_\rho}| \geq 1- \frac{1}{4\ell^{1-a}}.
\end{equation*}

\subsection{Proof of thermalization of the reference}
\label{sec:asymp_rigorous}

To check that the reference asymptotically reaches the thermal state discussed in the previous section, one can resort to a brute-force calculation that $\exp(-\beta L_z)$ is a fixed point of $\cE_\xi$. Another instructive way to see this begins by writing $\rho$ in the basis $\ket{m}$ which diagonalizes $L_z$.  By inspecting Eq.~\ref{alternateform}, we see that $\cE_\xi$ does not mix the various diagonals of $\rho$, i.e.\, 
\[\cE_\xi(\ket{m}\bra{m'}) = \sum_{a=-1}^1 \widetilde{P}_{(m+a,m'+a)|(m,m')}\ket{m\!+\!a}\bra{m'\!+\!a}\]
for nonnegative numbers $\widetilde{P}_{(m+a,m'+a)|(m,m')}$ satisfying 
\[\sum_{a=-1}^1\widetilde{P}_{(m+a,m'+a)|(m,m')} \leq 1.\]
One may further check that this bound is saturated only when $m=m'$.  Therefore, when $m\neq m'$, these weights correspond to subnormalized conditional probabilities which asymptotically annihilate the off-diagonal terms of $\rho$.  Abbreviating
\[P_{m+a|m} \equiv \widetilde{P}_{(m+a,m+a)|(m,m)},\]
we may express the action of $\cE_\xi$
on diagonal states (those which commute with $L_z$) as
\[\cE_\xi(\proj{m}) = \sum_{a=-1}^1
P_{m+a|m}\proj{m+a}.\]
The conditional probabilities can be directly evaluated using  Eq.~\ref{alternateform}, obtaining 
\begin{eqnarray}
P_{m\pm 1|m} &=& \frac{1 \pm 2\vev{S_z}}{d^2}\,\big|\bra{m\pm 1}L_\pm\ket{m}\big|^2 \nonumber \\
&=& \frac{1\pm 2\vev{S_z}}{4}\left(1 - \left(\frac{2m \pm 1}{d}\right)^2\right)
\label{eq:Markov}
\end{eqnarray}
and $P_{m|m} = 1-P_{m + 1|m} - P_{m-1|m}$.
This defines a Markov chain whose stationary state 
$\rho_\infty = \sum_m P_m \proj{ m}$ is found 
by demanding the detailed balance equation
\[P_mP_{m+1|m} = P_{m+1}P_{m|m+1},\]
which yields the unique solution
\[P_m \propto \left(\frac{1-2\vev{S_z}}{1+2\vev{S_z}}\right)^m = \exp(-\beta m).\] 
This proves that the asymptotic quantum partition function of the reference is indeed  $Z= \Tr\exp(-\beta L_z)$. 

\section{Thermalization time}
\label{sec:time}

So far, we have derived expressions for the asymptotic state of the reference and have characterized its ability to measure the spin of the source particles in this state. We will now analyze how much time it takes for the reference to reach this asymptotic state. In circumstances where the asymptotic state is not a useful resource --- i.e. when it becomes misaligned or fuzzy --- this thermalization time will set a fundamental upper bound on the longevity of the reference.  

\subsection{A lower bound on the thermalization time via conservation laws}
\label{sec:timeA}
Here, we give a straightforward derivation of a lower bound on the thermalization time, using only basic symmetry arguments. Since measuring the total angular momentum of the combined system is a rotationally invariant process, it preserves angular momentum. To significantly disturb the state of a reference with large angular momentum requires ``kicking it" with comparable amount of angular momentum. When each source particle is in a state with angular momentum $\vev{S_z}$, the total angular momentum of the source particles is a binomial random variable.  Therefore, the 
{\em typical} states of $t$ source particles will have a total angular momentum given by  
\begin{equation*}
\sum S_z\Big|_\text{typical} = t\vev{S_z} \pm \sqrt{{\textstyle \frac{1}{4}} - \vev{S_z}^2}\sqrt t.
\end{equation*} 

In the case of an unpolarized source ($|\vev{S_z}| < 1/\ell$), the physical requirement to disturb the state of the reference leads to a number of measured particles $t \gtrsim \ell^2$; the disturbance is due to \emph{fluctuations} of the source's angular momentum. This explains the main results of \cite{BRST06a,BRST06b}. On the other hand, when the source is sufficiently polarized ($|\vev{S_z}| > 1/\ell $), the disturbance is no longer dominated by fluctuations but rather by the bias, or drift, leading to a longevity $t \gtrsim \ell/2\vev{S_z}$.

This argument is not restricted to the particular $SU(2)$ model we have been studying here; it should extend to arbitrary reference systems as well. In particular, when the source particles are in random states, the expectation value of any traceless local operator will grow as the square-root of the number of particles. This yields a strict lower bound on the longevity of the reference.

\subsection{An upper bound on the thermalization time} 

We have demonstrated in Sec.~\ref{sec:asymp_rigorous} how the map $\cE_\xi$ reduces to a Markov chain on density operators diagonal in the $L_z$ basis. Thus, we can use standard tools to analyze the expected hitting time for any state. The expected hitting time of a state $\ket{\ell,m}$ is the average time it takes the system to reach this state. 

Consider the case where the reference begins in a coherent state which points in the $-\b{z}$ direction, i.e.\ it is initially in the pure state $\ket{\ell,-\!\ell}$, and the source is maximally polarized, i.e.\ $\vev{S_z} = \frac 12$. Intuitively, we expect this ``antiparallel" setting to be a worst case scenario, and so to yield a fundamental lower bound on the longevity of the reference.  

Substituting the value of $\vev{S_z}$ into Eq.~(\ref{eq:Markov}), we see that at every time step, the angular momentum of the reference will either increase by $\frac 12$, or will otherwise stay the same.  In this case, the unique steady state is $\ket{\ell,\ell}$ as expected. The average time $T$ required to arrive at this state starting from the state $\ket{\ell,-\ell}$ can be computed by adding the average time required to go from state $\ket{\ell,m}$ to $\ket{\ell,m+1}$ for $m=-\ell, -\ell+1, \ldots, \ell-1$:
\begin{eqnarray*}
T &=& \sum_{m=-\ell}^{\ell-1} \frac{2}{1-\left(\frac{2m+1}{d}\right)^2} \\
&\approx&  d\int_{-1+\epsilon}^{1-\epsilon} \frac{1}{1-y^2} dy \\
&=& d\log(d-1) 
\end{eqnarray*}
where $\epsilon = 2/d$.  If the reference is in the fixed point, it gives the incorrect result with certainty: this is not surprising since it started pointing in the $-\b{z}$ direction and ended up pointing in the $+\b{z}$ direction. Thus, $T$ gives an upper bound on the longevity of the reference. 

Note that the logarithmic factor is due to our convergence criterion becoming more strenuous as we increase $\ell$. Had we instead computed the average time to reach a state with $\vev{L_z}/\ell \geq 1-\epsilon$ starting in a state with $\vev{L_z}/\ell \leq 1-\epsilon$ where $\epsilon$ is a constant {\em independent} of $d$, we would have obtained a linear hitting time $T = d\log(\frac 2\epsilon -1)$. Since this matches the lower bound obtained in the previous section, we conclude that this scaling is asymptotically tight.  Hence, we conclude that the longevity of the reference in this worst case is upper bounded by a linear function of the size of the reference.

\begin{figure}[tbh!]
\center \includegraphics[width=2.6in]{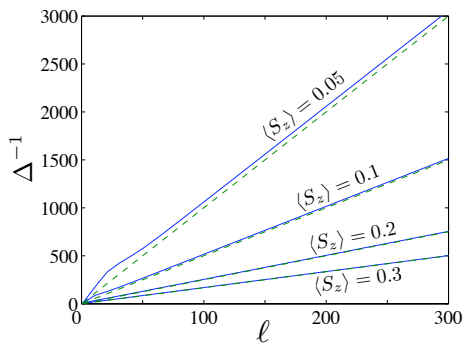}
\caption{Gap inverse ($\Delta^{-1}$) of the matrix $W$ as a function of the reference size $\ell$ and for different values of $\langle S_z \rangle  = 0.05, 0.1, 0.2,$ and $0.3$. Dashed lines indicates the estimate $\ell/(2\langle S_z \rangle)$.}
\label{fig:gap}
\end{figure}

For more general settings, this thermalizing time can be evaluated using the spectral gap theorem. Given the stochastic Markov matrix $P_{mm'}$ defined by Eq.~(\ref{eq:Markov}), we construct the matrix $W = \rho(\infty)^{-1/2}P \rho(\infty)^{1/2}$. The convergence time is equal to the inverse of the gap between the two largest eigenvalues of $W$. We have found numerically that this gap is given by $\Delta \approx 2\vev{S_z}/\ell$ as pictured in Fig.~\ref{fig:gap}. Thus, the time required to reach the fixed state is $t \propto \ell/2\vev{S_z}$, once again in agreement with the predictions of the previous section.

%\subsection{Implications for the longevity of the reference}

\section{Non-equilibrium state}
\label{sec:non-eq.}

\begin{figure}[tbh!]
\center \includegraphics[width=2.8in]{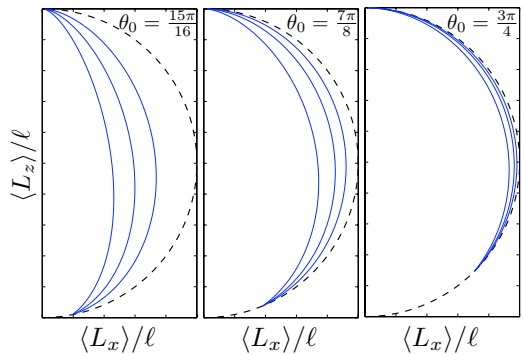}
\caption{Phase space trajectories $(\langle L_x(t)\rangle /\ell,\langle L_z(t)\rangle /\ell)$ of the equilibration process for different initial states $\ket{\theta_0}$, and different reference sizes $\ell=20$, 40, and 80. The dashed line represents the manifold of coherent states, for which $\langle L_x\rangle^2 + \langle L_z\rangle^2 = \ell^2$. The source is maximally polarized $\langle S_z\rangle = 1/2$. The loss of polarization of the reference during equilibration is suppressed as its size increases. In the macroscopic limit, the reference remains in a coherent state throughout the process, except for the singular initial state $\theta_0 = \pi$.}
\label{fig:phase}
\end{figure}

We have studied the asymptotic state of the reference and characterized the time it takes to get there. Our final task is to study the thermalization process itself --- to describe the dynamics of the reference as it reaches its equilibrium state. More specifically, since the expectation $\vev{{\bf L}}/\ell$ fully characterizes the operational behavior of the reference, our aim is to derive an equation for the time evolution of this quantity. As we did before, we choose our coordinates so that the source's polarization is along the $z$ axis and the initial orientation of the reference is at an angle $\theta$ from the $z$ axis in the $z-x$ plane, or in other words $\vev{L_y} = 0$. By symmetry, we know that $\vev{L_y}$ will remain 0 for all times. Thus, we can parameterize $\bf n$ by its length $r$ and its angle $\theta$ away from the $z$ axis. Figure~\ref{fig:phase} displays the thermalization trajectories of the reference for different reference sizes $\ell$ and different initial coherent states $\ket{\theta_0} = \exp\{-i\theta_0 L_y\}\ket{\ell,\ell}$. 

To derive an equation of motion, it is convenient to express the CPTP map $\cE_\xi$ from Eq.~(\ref{eq:channel1}) acting on the reference using a rotated basis of operators 
$\b{L}^\theta$, given by \[(L_x^\theta,L_y^\theta,L_z^\theta) = (\cos\theta L_x - \sin \theta L_z, L_y,\sin\theta L_x + \cos\theta L_z).\] 
In particular, we choose $\theta$ so that $\vev{L_x^\theta} = \vev{L_y^\theta} = 0$ and $\vev{L_z^\theta} = r\ell$ for some $0 \leq r \leq 1$. With some straightforward manipulations, we can express this map in the Kraus form $\cE_\xi(\rho) = \sum_a E_a\rho E_a^\dagger$ where the four Kraus operators $E_a$ are given by 
\begin{eqnarray*}
\frac{1}{d\sqrt 2}\Big\{&&
\hspace{-.15in}\sqrt{d^2+1-4\vev{S_z}^2}, \ 
 i2\sqrt{1-4\vev{S_z}^2} L_y^\theta \\
&& \hspace{-.13in} 2L_z^\theta + i4 \vev{S_z} \sin\theta L_y^\theta + 2\vev{S_z}\cos\theta,\\
&& \hspace{-.13in}2L_x^\theta + i4\vev{S_z}\cos \theta L_y^\theta - 2\vev{S_z}\sin\theta
%&& (1+2\vev{S_z}\cos\theta)L_+^\theta + (1-2\vev{S_z}\cos\theta)L_-^\theta \\
%&& \,\,-\, 2\vev{S_z}\sin\theta
\Big\}.
\end{eqnarray*}

\begin{figure}[tbh!]
\center \includegraphics[width=2.5in]{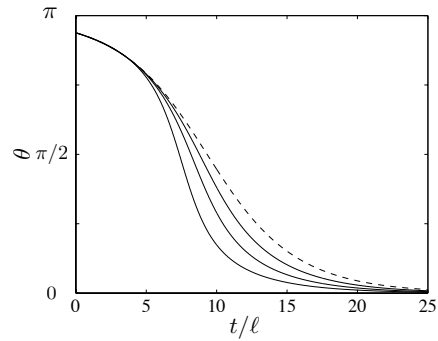}
\caption{Orientation of the polarization of the reference relative to the $z$ axis as a function of time. The source's polarization is non maximal $\vev{S_z} = 1/4$ and the reference is initially in the coherent state $\theta_0 = 15\pi/16$. The dotted line is the prediction obtained from Eq.~(\ref{eq:diff_theta}).  Solid lines represent the exact numerical solutions for $\ell = 20$, 40, and 80.  As the size of the reference increases, the exact trajectories approach the semi-classical solution $\theta_{SC}(t)$.}
\label{fig:theta}
\end{figure}

To make further progress, observe that when acting on the state of the reference, $L^\theta_z \sim \ell r$ and $L_x^\theta, L_y^\theta \sim \sqrt{\ell}$. Expanding in powers of $1/\ell$ and keeping only first-order terms, we obtain the approximation
\begin{equation}
\cE_\xi(\rho) \approx
\rho + i\frac{r\vev{S_z}}{\ell}\sin\theta[L_y,\rho].
\end{equation} 
Observe that the second term induces a rotation by an angle $rz\sin\theta/\ell$ about the $y$ axis.  This approximation leads to the following equation for the angle $\theta(t)$ relative to the $z$ axis
\begin{equation}
\frac{d\theta}{dt} = -r(t)\frac{\vev{S_z}}{\ell}\sin\theta +O(1/\ell^2).
\label{eq:diff_theta}
\end{equation}
By numerical analysis which is summarized in Fig.~\ref{fig:phase}, we see that in the macroscopic limit, the polarization of the reference remains close to $1$ throughout the entire trajectory, provided that the initial polarization of the reference is sufficiently far from being anti-parallel with that of the source.  Assuming that $r(t) =1$ throughout the entire trajectory, we may thus solve Eq.~(\ref{eq:diff_theta}) to obtain a semi-classical trajectory: \[\theta_{SC}(t) = 2\mathrm{arccot}\big(\mathrm{cot}(\theta_0/2)e^{\vev{S_z}t/\ell}\big).\] 
%where $a_0 = \mathrm{cot}(\theta_0/2)$. 
This solution is compared with the exact numerical evaluation for various reference sizes in Fig.~\ref{fig:theta}. We note that this derivation justifies the {\em ad hoc} semi-classical model used in \cite{BRST06b}.

\section{Discussion}
\label{sec:discussion}

We have analyzed the behavior of a quantum gyroscope which is modeled as a large spin and used to measure sequences of spin-$\frac 12$ source particles from a fixed ensemble.  We showed that the measurement which is actually performed on the source depends only on the angular momentum of the reference.  We then derived expressions for the quantum channel which describes the back-action on the reference and showed that its unique fixed point is a thermal state.  
It is also possible to characterize the asymptotic state of the reference as  
\[\rho(\infty) \propto \Pi_\sym \xi^{\otimes 2\ell} \Pi_\sym,\] where $\Pi_\sym$ is the projector onto the ($2\ell + 1$ dimensional) symmetric subspace of $(\bC^2)^{\otimes 2\ell}$. This expression is insightful when the reference is a composite system made up of $2\ell$ spin-$\frac 12$ particles, as one can imagine that the particles of the reference get ``exchanged" with those of the source, while remaining in the spin-$\ell$ representation.  

We then estimated the amount of time it takes to approach the thermal state.  For a polarized source, we saw that the thermalization time scales \emph{linearly} with the size of the reference.  The implications of this for the longevity of the reference are two-fold.  If the source and reference begin in alignment, the longevity will be \emph{infinite} for a sufficiently large (but still finite) reference size.  This is because the asymptotic state will still have most of its angular momentum pointing in the original direction.  In all other cases with a polarized source, the reference will reorient itself to become aligned with the source after a time that scales linearly with its size. 

We also gave a physical mechanism describing the  main result from \cite{BRST06a}: for an unpolarized source, the thermalization time scales \emph{quadratically} with the reference size because the dynamics is dominated by fluctuations rather than by drift.  Therefore, its longevity enjoys the same scaling.  However, we saw that this type of behavior is quite non-generic; unless the angular momentum of the source is \emph{exactly} zero, a sufficiently large (but finite) reference will asymptotically retain a near maximal amount of angular momentum.  The quadratic scaling, on the other hand, requires a precisely tuned source, reminiscent of the effort needed to ensure that a statistical system remains in the vicinity of its critical point.

 Finally, we initiated the analysis of the non-equilibrium dynamics of the reference, giving a semiclassical equation for the time-evolution of the reference.  In particular, we found that if the reference begins in a coherent state, it will essentially remain in a coherent state throughout its evolution, provided that the source particles are not anti-parallel to the initial state of the reference.   Here, the reference simply rotates itself toward the final state in an essentially \emph{deterministic} fashion.  Thus, while it loses its ability to perform the measurement for which is was originally intended, it will always be reasonably good at performing \emph{some} particular measurement (which is predicted by our equation) at any given time.  For an even better understanding of the semi-classical dynamics, it would be beneficial to also have an equation for the time evolution of the length $r$ of the reference, or at least to obtain an analytic bound on its minimal value throughout its evolution.    
 
As mentioned in the introduction, our analysis is relevant for the realization of quantum information processors. For micro-fabricated architectures, some components of the measurement device can be subject to quantum fluctuations and will thus feature effects similar to those analyzed in this paper. Most schemes for fault-tolerant quantum computation (see e.g.\ \cite{Sho96a}) utilize repeated measurements for syndrome extraction and post-selected state distillation. Moreover, the measured particles will typically exhibit a strong bias (e.g.\ syndromes are more likely to be 0 than 1), inducing a rapid degradation of the reference.  On the other hand, it may also be possible to utilize such a source bias for stabilizing a reference, allowing sufficiently accurate measurements to be performed throughout an arbitrarily long computation. 
 
One could imagine extending the analysis put forth in this paper in a number of ways.  For example, the reference could be used to measure the angular momentum of particles in higher representations of $SU(2)$.  Additionally, more complicated Lie groups could be substituted; repeating our analysis in such cases would involve even further generalizations of channels studied in \cite{Rit05a}. As many of our arguments are based on general physical principles, we expect that the same scaling properties derived here would apply in those cases as well.  

\section{Acknowledgements}
We thank Stephen Bartlett and Peter Turner for comments on an earlier version of this manuscript, and Florian Girelli and Wojciech Zurek for useful conversations. This work was supported in part by the National Science Foundation under Grant No. PHY-0456720. DP also receives financial support from the Gordon and Betty Moore Foundation through Caltech's Center for the Physics of Information and from the Natural Sciences and Engineering Research Council of Canada. 

%\bibliographystyle{/Users/dpoulin/archive/qubib}
%\bibliography{/Users/dpoulin/archive/qubib}

\newcommand{\etalchar}[1]{$^{#1}$}

\end{document}